\documentclass[%
reprint,
 amsmath,amssymb,
 aps,
prb,
]{revtex4-1}

\usepackage{color}
\usepackage{graphicx}
\usepackage{dcolumn}
\usepackage{bm}

\begin{document}

\title{{\it Ab initio}\ typical medium theory of substitutional disorder}

\author{A. \"Ostlin$^{a}$,
Y.\ Zhang$^{b,c}$, H. Terletska$^d$,
F. Beiu\c seanu$^e$,
V. Popescu$^f$ \\
K. Byczuk$^g$, L. Vitos$^{h,i,j}$,
M. Jarrell$^{b}$,
D. Vollhardt$^a$, L. Chioncel$^{a,k}$}

\affiliation{$^a$Theoretical Physics III, Center for Electronic Correlations and Magnetism, Institute of Physics, University of Augsburg, D-86135 Augsburg, Germany}
\affiliation{$^b$Department of Physics \& Astronomy and Center for Computation \& Technology, Louisiana State University, Baton Rouge, Louisiana 70803, USA}
\affiliation{$^c$Kavli Institute for Theoretical Sciences, Beijing 100190, China}
\affiliation{$^d$Department of Physics and Astronomy, Middle Tennessee State University, Murfreesboro, Tennessee 37132, USA}
\affiliation{$^e$Faculty of Science, University of Oradea, RO-410087 Oradea, Romania}
\affiliation{$^f$Sophie-Scholl-Gymnasium Oberhausen, 46145 Oberhausen,
Germany}
\affiliation{$^g$Institute of Theoretical Physics, Faculty of Physics, University of Warsaw, ul. Pasteura 5, PL-02-093 Warszawa, Poland}
\affiliation{$^h$Department of Materials Science and Engineering, Applied Materials Physics,
KTH Royal Institute of Technology, SE-10044 Stockholm, Sweden}
\affiliation{$^i$Department of Physics and Materials Science, Division for Materials Theory, Uppsala University, SE-75121 Uppsala, Sweden}
\affiliation{$^j$Research Institute for Solid State Physics and Optics, Hungarian Academy of Sciences, P.O. Box 49, H-1525 Budapest, Hungary}
\affiliation{$^k$Augsburg Center for Innovative Technologies, University of Augsburg, D-86135 Augsburg, Germany}

\date{\today}

\begin{abstract}
    By merging single-site typical medium theory with density functional theory we introduce a self-consistent framework for electronic structure calculations of  materials with substitutional disorder which takes into account Anderson localization. The scheme and details of the implementation are presented and applied to the  hypothetical alloy Li$_{c}$Be$_{1-c}$, and the results are compared with those obtained with the coherent potential approximation. Furthermore we demonstrate that Anderson localization suppresses ferromagnetic order for a very low concentration of
    (i)  carbon impurities substituting oxygen in MgO$_{1-c}$C$_{c}$, and
    (ii) manganese impurities substituting magnesium in Mg$_{1-c}$Mn$_c$O for the low-spin magnetic configuration.
\end{abstract}


\maketitle

\section{Introduction}
The unusual electronic properties of disordered metals and
alloys ~\cite{ziman.diso,le.ra.85,be.ki.94} result from the absence of translational invariance in such systems.
To calculate the physical properties of disordered solids and, in particular,  to determine their
electronic structure is still a challenging
problem which involves the sampling of random configurations followed by a quantum-mechanical
computation for each disorder configuration. Disordered systems are usually modeled numerically by large supercells in real space, where the results must then
be averaged over different realizations of the disorder.
This  increases the cost of electronic structure calculations of disordered systems considerably.

In this paper we introduce a computational approach which merges  effective--medium
approximations with the Density-Functional Theory~\cite{ho.ko.64,ko.sh.65,kohn.99,jo.gu.89,jone.15} (DFT)  to investigate structurally disordered solids.
The effective medium is calculated by means of a statistical approach,
which takes into account the strength of the disordered
alloy potential. The latter can be decomposed into a sum
of contributions from the individual atomic scatterers, such that
the electron propagation can be viewed as a succession of scatterings from these atomic potentials. The essence of the effective--medium theory
is the self-consistent treatment of the multiple-scattering events: the scatterers are viewed as
embedded in an effective medium whose properties still have to be determined.
If the average
scattering from a single impurity in the presence of the effective medium is set to zero one obtains
the  well-known Coherent Potential Approximation (CPA) \cite{el.kr.74,ziman.diso}.
The CPA was introduced by
Soven~\cite{sove.67} and
Taylor~\cite{tayl.67} to study electronic and vibrational properties of random alloys, respectively. It was then further developed  and extensively applied to disordered solids~\cite{ve.ki.68,el.kr.74,ziman.diso,le.ra.85,be.ki.94}.
To investigate the electronic structure of realistic materials Gy\H{o}rffy \cite{gyor.72}
formulated the CPA in the framework of multiple scattering theory by
using the Green's function technique.

Even today the CPA is one of the most widely employed methods to calculate the
electronic structure of random alloys. Numerous
applications \cite{vi.ko.02,fa.st.80,fa.be.82,jo.ni.86,%
we.dr.94,si.go.93,jo.ni.90,%
ma.gr.02,ma.gr.02b,os.vi.18} have shown that within this
approximation one can calculate lattice parameter, bulk modulus,
mixing enthalpy, etc., with an accuracy similar to that
obtained for ordered solids.
At the same time the applicability of the CPA is limited since it is a
single-site approximation. For example, CPA neither takes into account disorder--induced
short-range  correlations, nor the effects of Anderson localization\cite{le.ra.85}. Furthermore, systems with a
large size mismatch between the alloy components are difficult to treat within CPA
because of the local lattice relaxations.
The search for
a generalization of the CPA has proven to be difficult.
There have been numerous attempts to overcome the main
shortcomings of the CPA by incorporating the missing nonlocal
physics, e.g., by employing the molecular CPA~\cite{go.ga.78,goni.92} and the
dynamical cluster approximation (DCA)~\cite{ja.kr.01} in model
Hamiltonian calculations, and the KKR-Non-Local Coherent Potential
Approximation (KKR-NLCPA)~\cite{ro.st.03,bi.gh.05,ro.st.05,ro.er.06}, where the DCA
coarse graining approach is applied.

The purpose of the present paper is to introduce an alternative
choice of the effective medium, which is able to take into account the effects of Anderson localization even in a real disordered material, by
formulating it within the framework of DFT. As pointed out by Anderson~\cite{ande.58}, the key quantity  to study  in a disordered system is the
amplitude
of the electronic wave functions. At the  localization transition the
 Hamiltonian spectrum in the vicinity of a given energy
changes from continuous to discrete (dense-point-like) in the thermodynamic limit and the
{\it typical},  i.e., the most probable value of the local density of states (LDOS)  at
this energy, vanishes.  The typical value  of the LDOS  is well represented by the
geometric average~\footnote{For discussions see~\cite{do.pa.03,by.ho.05,by.ho.10}}
$\rho_{typ} (E)= \exp[\langle \ln \ \rho_i (E) \rangle]$,
where $\langle ... \rangle$ represents the arithmetic average over disorder  and $\rho_i (E) $ is the LDOS at site $i$ for the energy $E$~\cite{sc.sc.10}.
This technique was successfully  applied  to model Hamiltonians
and is referred to as ``typical medium theory'' (TMT)~\cite{do.pa.03,dobr.10,ma.ta.15, ag.do.09, ag.do.13, br.ag.15,ag.do.06}.
The TMT can describe signatures of Anderson localization in the spectral function, i.e., on the one-particle level, but does not capture short-range order effects.

Recently, the typical medium dynamical cluster approximation
(TMDCA)~\cite{ek.te.14,te.zh.18} was introduced, which extends the single-site TMT to a finite cluster
and allows for a systematic inclusion of the nonlocal multisite correlations.
It was shown that the TMDCA overcomes many shortcomings of the TMT since it allows one to identify effects of Anderson localization in higher-order correlation functions, e.g., the conductivity.
This method has also been extended to models with off-diagonal
disorder~\cite{te.ek.14}, multiband systems~\cite{yi.te.15}, and
interactions~\cite{ek.ya.15}.
To go beyond model studies and ultimately investigate also real materials the TMDCA was subsequently
formulated within the framework of multiple scattering theory~\cite{te.zh.17}.
However, so far this framework was only applied to model
Hamiltonians~\cite{ek.te.14,yi.te.15,te.zh.18}.
While some of these models were extracted from first principles
calculations~\cite{yi.te.15,zh.ne.16,zh.ne.18}, no
self-consistent feed-back between the models and the typical
medium analysis was considered.

The aim of our paper is to extend this methodology from
 models to realistic three-dimensional muffin-tin systems by
merging the single-site typical medium theory with DFT.

The paper is organized as follows.  In Sec.~\ref{effectiv_med} we introduce two effective medium theories
which have been formulated to compute electronic structures using Green's functions.
In Sec.~\ref{EMTO_effectiv_medium} we review the theory using
the exact muffin-tin orbitals (EMTO) basis set and present the form of the Green's function and
path operators. The CPA- and TMT-Green's function condition for the self-consistency of the effective medium are
discussed in Sec.~\ref{sec:eff_med}.
The method is then applied to compute the
electronic properties of the hypothetical Li$_{c}$Be$_{1-c}$
alloy and the dilute MgO$_{1-c}$C$_c$ and Mg$_{1-c}$Mn$_c$O alloys.
The Li-Be system is used as an illustrative example to compare
CPA vs. TMT calculations, and to discuss signatures of the precursor
of the Anderson localization transition. In the dilute MgO$_{1-c}$C$_{c}$
and Mg$_{1-c}$Mn$_c$O alloys we investigate how magnetism is
influenced by weak disorder (within CPA) and strong disorder
(within TMT), respectively.

\section{Single-site effective-medium theories}
\label{effectiv_med}
 We begin with a general description of the effective-medium theory of realistic multi-atom alloys. To this end we consider a substitutional alloy A$_a$B$_b$C$_c$..., where the
atoms A, B, C, ... are randomly distributed on the underlying
crystal structure. Here $a$, $b$, $c, ...$ stand for the atomic
fractions (concentrations) of the A, B, C, ... atoms, respectively.
 In the alloy it is assumed that the potential of the atoms at any lattice site
 is completely random and  can be described by a probability distribution function (PDF)
 $P(\epsilon_1,...\epsilon_N) $ for local energy levels $\epsilon_i$, where $N$ is the total number of atoms in the sample.
This allows us to determine the expectation
value of an observable ${\mathcal A}(\epsilon_1,...,\epsilon_N)$ as the arithmetic  average over
different disorder realizations with
this PDF, i.e.,
$\tilde{{\mathcal A}} = \langle {\mathcal A} \rangle_{\mbox{\tiny arith}} =
\int_{-\infty}^{\infty}  \prod_{i=1}^N d\epsilon_i {\mathcal A}(\epsilon_1,...,\epsilon_N) P(\epsilon_1,...,\epsilon_N)$.

There are two major
approximations in the single-site effective medium construction:
(i) the local
potentials around one type of atoms forming the alloy are assumed to be the
same, i.e., the effect of the local environment is neglected.
 Therefore, the PDF is uncorrelated and has a product form, where the PDF for each type of atom is given by
\footnote{For the definition of the PDF function
within the muffin-tin formalism see~\cite{ande.75,skr.84}}
$P_{\rm A}, P_{\rm B}, P_{\rm C},...$, respectively;
(ii) The system is replaced by a mono-atomic  effective crystal  described by
the site-independent ``effective medium potential'' ${\tilde D}$.
One therefore approximates the Green's function $g$ of a real system by an ``effective medium
Green's function'' ${\tilde g}$, and for each alloy component $j=$A, B,
C,... a single-site Green's function $g_j$ is determined.
The construction of such a single-site effective
medium involves the following steps:
\begin{itemize}
\item[1)] The effective Green's function is calculated from the
effective potential using an electronic structure method.
For example, within the Korringa-Kohn-Rostoker (KKR)
\cite{korr.47,ko.ro.54,wein.90,goni.92,ha.se.61,fa.da.67} or
Linear Muffin-Tin Orbital (LMTO)
\cite{ande.75,skr.84,ande.75} methods, one has
\begin{equation}\label{eff1}
{\tilde g} = \left[S - {\tilde D}\right]^{-1},
\end{equation}
where $S$ denotes the KKR or LMTO structure constant matrix
corresponding to the underlying lattice.
\item[2)] Next, the Green's functions $g_j$
of the alloy components  are determined by substituting  the real atomic potentials
$D_j$ by their value computed with respect to the effective medium potential $\tilde{D}$. Mathematically, this condition is expressed by the
real-space Dyson equation
\begin{eqnarray}\label{eff2}
g_j = {\tilde g} + {\tilde g}\left (D_j - {\tilde
D}\right)g_j\;\;,\;\;j={\rm A,B,C}...
\end{eqnarray}
\item[3)] Finally, the average of the individual Green's functions should
reproduce the single-site part of the
effective medium Green's function,
i.e.,
\begin{equation}\label{eff3}
{\tilde g} = {\tilde g}  \left[ \ g_{\rm A},  g_{\rm B}, g_{\rm C} , ...\right].
\end{equation}
This functional relation needs to be specified for each type of effective medium theory.
\end{itemize}

Eqs. (\ref{eff1})$-$(\ref{eff3}) are solved iteratively, and
the output  functions ${\tilde g}$ and $g_j$ are used to determine the
electronic structure, charge density, and the total energy of the random
alloy.

If $\tilde{g}$ is determined by the arithmetic average the results are equivalent
to the CPA and are insensitive to Anderson localization. By contrast, the geometric
average leads to the TMT which is capable of describing disorder-driven Anderson
localization effects. In the next Section we will present this construction for the TMT  explicitly.

\section{Effective medium theory
 using exact muffin-tin orbitals}
\label{EMTO_effectiv_medium}

The EMTO theory~\cite{an.sa.00,vito.01} formulates an efficient and at the same time
accurate muffin-tin method for solving the Kohn-Sham equations~\cite{ho.ko.64,ko.sh.65}
of the DFT~\cite{ho.ko.64,ko.sh.65,kohn.99,jo.gu.89,jone.15}. By using large overlapping potential
spheres, the EMTO approach describes the exact
crystal potential  more accurately than any other conventional muffin-tin method.
In the EMTO approach, while keeping the simplicity and
efficiency of the muffin-tin formalism, the one-electron states are  determined
exactly for the model potential.

\subsection{Effective medium potential}

Within the overlapping muffin-tin approximation the Kohn-Sham effective
potential $v(\textbf{r})$ for a real alloy is approximated by spherical potential wells $v_R^j(r_R)$
centered on atomic sites $\textbf{R}$, where the subscript $j$ denotes the alloy component at site $\textbf{R}$, supplemented by a constant interstitial potential
$v_0$, i.e.
\begin{equation}
v(\textbf{r}) \approx v_{\rm mt}(r) \equiv v_0 + \sum\limits_{R}
[v^j_R(r_R)-v_0].
\label{potential}
\end{equation}
Here we introduced the notation $\textbf{r}_R \equiv  \textbf{r}-\textbf{R} \equiv  r_R \hat{r}_R$, where $\hat{r}_R=\textbf{r}_R/|\textbf{r}_R|$, and $r_R=|\textbf{r}_R|$.
In the following we will omit the explicit vector notation for simplicity.

Next we consider a substitutional alloy with a fixed underlying lattice.
We denote the  positions of the atoms
of the underlying lattice by $R$,
$R^\prime$, etc. in a given unit cell.
Within the unit cell we  can have one of the atoms from the $N_R$ components forming an alloy,
 e.g., for a binary-alloy $N_R=2$.
The atomic fractions of the components determine the concentrations
$c^j_R$ ($j=1,2,...,N_R$).
The individual spherical potentials in Eq.~(\ref{potential}),
denoted by $v^j_R(r_R)$,  are defined within the potential
spheres of radii $s_R^j$. We note that these potentials are
not exactly the same as the spherical potentials  present in
a real alloy because of different local environments.
Within effective medium theories we assume
that all potential-dependent functions, such as the
partial waves, logarithmic derivatives, normalization functions,
etc., belonging to the same kind  of atom within a given unit cell,
are the same.

The effective medium is described by a site-dependent
effective potential (subscript $R$), which possesses the symmetry of the underlying
crystal lattice.

\subsection{Effective medium Green's function}
\label{sec:eff_med}

In the EMTO formalism, the effective potential is
introduced via the logarithmic derivative ${\tilde D}_{RL'RL}(z)$ of
the effective scatterers. Therefore the coherent Green's
function or the path operator (\ref{eff1}) is given by
\cite{vito.01}:
\begin{eqnarray}\label{g(z)}
\sum_{R''L''}&\;&a_{Q'}\left[S_{R'L'R''L''}(\kappa^2,{\bf k})\;-\;
\delta_{R'R''} {\tilde
D}_{R'L'R'L''}(z)\right]\nonumber\\&\times&{\tilde
g}_{R''L''RL}(z,{\bf k})\;= \;\delta_{R'R} \delta_{L'L},
\end{eqnarray}
where $l,l',l'' \le l_{max}$, and $S_{R'L'R''L''}(\kappa^2,{\bf k})$
are the elements of the EMTO slope matrix for complex energy
$\kappa^2=z-v_0$ and Bloch vector ${\bf k}$ from the Brillouin zone
(BZ). The logarithmic derivative of the effective
scatterers is site-diagonal with non-zero $L'\ne L$ off-diagonal
elements.

The local part of the Green's function of the alloy component
 $g^j_{RL,RL^\prime}$ is calculated as an impurity
Green's function embedded in the
effective medium. In the single-site approximation this is obtained
from the real space Dyson equation (\ref{eff2}) as a single-site
perturbation on the coherent potential as
\begin{eqnarray}\label{gi}
g^j_{RLRL'}(z)\;=\;{\tilde g}_{RLRL'}(z)\;+\;
\sum_{L''L'''}{\tilde
g}_{RLRL''}(z)\nonumber\\ \times\left[D^j_{Rl''}(z)
\delta_{L''L'''}\;-\;{\tilde
D}_{RL''RL'''}(z)\right]g_{RL'''RL'}^j(z).
\end{eqnarray}
Here $D^j_{Rl}(z)$ is the logarithmic derivative
function for the $j$th alloy component and
${\tilde g}_{RLRL'}(z)\;=\;\int_{BZ} {\tilde g}_{RL,RL^\prime}(z,{\bf
k})\mathrm{d}{\bf k}$
is the site-diagonal part of the {\bf k}-integrated effective medium Green's
function. The condition of zero average scattering
leads to a functional relation
between ${\tilde g}_{RLRL'}(z)$ and
the Green's functions of alloy components,
namely
\begin{equation}\label{ggi}
{\tilde g}_{RLRL'}(z)\;=\; {\tilde g} \left[  g^j_{RLRL'}(z) \right],
\end{equation}
which still needs to be specified for different types of effective medium theories (see below).
Equations (\ref{g(z)}), (\ref{gi}), and (\ref{ggi}) are solved
self-consistently for ${\tilde D}(z)$, ${\tilde g}(z,{\bf k})$, and
$g^j(z)$. Note that the Green's functions for the alloy component and effective medium denoted as $g^j$ and $\tilde{g}$, respectively, carry the information
about the poles. Using the partial waves the Green's functions
are properly normalized.

As discussed earlier, there are two possibilities for determining properties of the effective medium:

\begin{itemize}
    \item {\it The Coherent potential condition} provides an expression for the coherent path operator $\tilde{g}$ as the algebraic average of the alloy-component path operators $g^j$ for an arbitrary complex argument $z$, i.e.,
 \begin{equation}\label{g_cpa}
{\tilde g}_{RLRL'}(z) :=       g_{RL,RL'}^{CPA}(z)
       = \sum_j c_R^j \; g^j_{RL,RL'}(z).
  \end{equation}
Obviously, the same algebraic condition is simultaneously fulfilled by the real and imaginary part of this coherent path operator.

    \item By contrast the {\it typical medium condition}
 is formulated along the
    real energy axis, i.e., the geometrical average is given by
\begin{equation}\label{rho_tm2}
 \rho_{RL}^{TMT}(E) := \prod_{j} \left[ \rho_{RL}^j(E) \right]^{c_R^j}.
 \end{equation}
Since the LDOS is defined
only for real energies $E$ it has to be computed at each iteration of the
 typical medium self-consistency loop. This makes it necessary to perform  analytic continuations between the complex-plane point $z$ and the real-axis point $E$.
To avoid the
 repetitive and CPU time consuming analytic continuation,  we perform  the geometric averaging in the current implementation
in the complex-plane and define the path operator as
\begin{equation}
      g_{RL,RL}^{TMT}(z) := \prod_j  \left[ g^j_{RL,RL}(z) \right]^{c_R^j}    \label{TM-g2} .
\end{equation}
Along the real axis Eq.~(\ref{TM-g2}) multiplied with the normalization of the partial waves (which is real on the real axis)
provides the LDOS $\rho_{RL}^j(E)$.
 We note that in general the condition expressed by Eq.~(\ref{TM-g2}) is an approximation to the TMT condition (9).
\end{itemize}

With this prescription of how to determine the effective medium in terms of
the path operator, the expression for the LDOS for a specific impurity $j$ at site $R$ is determined from
\begin{widetext}
\begin{eqnarray}\label{noscpa}
{\rho}^{j}_{RL}(z)  \;&\equiv&\; -\frac{1}{\pi} \Im  \bigg\{  \int_{BZ}\sum_{R'L'}{\tilde g}_{RLR'L'}(z,{\bf
k})\;a_{R'}{\dot S}_{R'L'RL}
(\kappa^2,{\bf k})d{\bf k} - g^j_{RLRL}(z) \dot{D}^j_{Rl}(z) - {\mathcal G}_{Rl}^{j,p}(z) \bigg\} ,
\end{eqnarray}
\end{widetext}
where  the over-dot denotes the energy derivative.
The off-diagonal ($R\ne R'$) elements of the coherent
Green's function ${\tilde g}_{R'L'RL}(z,{\bf k})$ are calculated from
Eq.~(\ref{g(z)}) with the self-consistently determined logarithmic
derivative ${\tilde D}_{RL'RL}(z)$ of the effective scatterers.
The first and second term on the right-hand side of Eq.~(\ref{noscpa})
assure the proper normalization of the one-electron states for the
optimized overlapping potential.
The third term in Eq.~(\ref{noscpa}) removes the unphysical pole of the logarithmic derivative.
This is a specific step in the EMTO implementation~\cite{vito.01}.
Altogether the three terms provide the EMTO Green's function of the effective medium.
The search for the Fermi level is similar to the standard CPA procedure as implemented in the EMTO~\cite{vito.01}.

\section{Applications}
\label{RESULTS}
We now apply the CPA-EMTO and TMT-EMTO theories to the  hypothetical alloy
Li$_{c}$Be$_{1-c}$ as well as to MgO$_{1-c}$C$_c$ and Mg$_{1-c}$Mn$_c$O in the
dilute limit (low concentrations $c$) and compare the results. The motivation to
study the LiBe system comes from the fact that it represents
a simple cubic system with one-atom per unit cell, for which the different effective-medium
averaging procedures can be tested readily.
At the same time it also allows one to identify
features of the LDOS which can be traced back to effects
of Anderson localization.
Our choice of MgO is motivated by the relatively large insulating gap.
Spin-polarized
in-gap states are formed upon carbon and manganese substitutions in MgO, an
effect that is sometimes linked to the so-called $d^0$ magnetism~\cite{el.ru.07,pa.pi.08}. We will show that such an effect can
be reproduced within the CPA. However, self-consistent TMT calculations provide
a different picture, namely that fluctuations in the number of particles
influence the magnetic stability. In addition to localization the spin
polarization is significantly reduced.

\subsection{Li$_{c}$Be$_{1-c}$ alloys: an illustrative example}

In pure Li and Be the atomic $s$-orbitals form a single band.
In Be, by partial substitution with Li, the local energies
become statistically independent (random) variables with
a distribution function
$$P(\epsilon_i)=c_{Li}\delta(\epsilon_i - \epsilon_{Li}) + c_{Be}\delta(\epsilon_i - \epsilon_{Be}),$$
where $c_{Li}$ and $c_{Be}$ are the concentration of Li and Be in the
alloy, respectively. The bandwidth is determined by the magnitude of the intersite
hopping and the coordination number. Anderson~\cite{ande.58} discovered
that when the width $\Gamma$ of the distribution $P(\epsilon_i)$ is
smaller than a critical value $\Gamma < \Gamma_c$, states in the
middle of the band are extended, while for $\Gamma \ge \Gamma_c$
all states in the band are localized. It has been shown~\cite{lifs.64,ha.la.66,zi.la.66,ha.la.67} that localized states can also exist in the tails of the DOS of disordered materials. Based on these arguments
Mott~\cite{mott.67} conjectured the existence of a
threshold energy (``mobility edge''), at which a sharp transition from localized to extended states occurs.

\begin{figure}[h]
\includegraphics[width=0.95\columnwidth,clip=true]{fig1a.eps}
\includegraphics[width=0.95\columnwidth,clip=true]{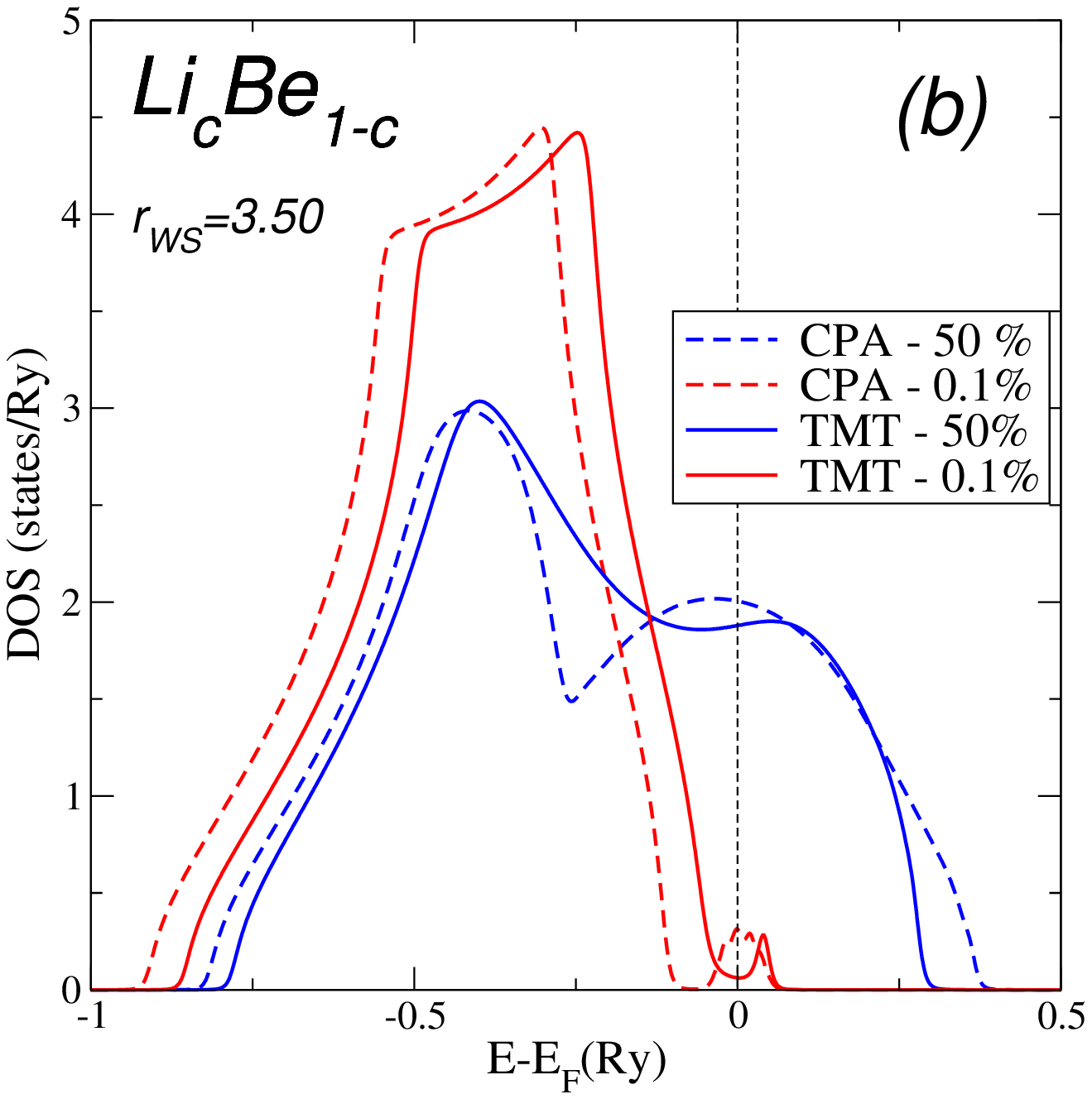}
\caption{(a) Average DOS
of Li$_{c}$Be$_{1-c}$ alloys calculated within CPA
for different concentrations $c$ for a given atomic Wigner-Seitz
radius $r_{WS}=3.50$~a.u., corresponding to a lattice parameter of
$5.64 \ a.u.$ ($=2.98$~\AA). (b) Average DOS
computed from the typical medium alloy components used in the charge self-consistent
calculations.} \label{cpa350}
\end{figure}

In Fig.~\ref{cpa350} we show the results for the
average DOS $\rho(E)=c_{Li}\rho^X_{Li}+c_{Be}\rho^X_{Be}$ which is computed from the alloy
components $\rho_{Li}$ and $\rho_{Be}$, where $X$=CPA or TMT.
Fig.~\ref{cpa350}~(a) corresponds to the DOS obtained for $X$=CPA,
where the alloy components are obtained from the CPA effective medium.
The concentration varies in the range from $1\%$ to $50\%$, and the
bandwidth corresponds to a fixed value of the Wigner-Seitz radius $r_{WS}=3.50$~a.u.
In the binary substitutional alloy  Li$_{c}$Be$_{1-c}$ the
Li-component has a reduced weight in comparison to the Be-component
for concentrations $c\le 0.5$.
In Fig.~\ref{cpa350}~(b) we compare the DOS obtained for $X$=CPA and $X$=TMT.

The general structure of the DOS shown in Fig.~\ref{cpa350} consists
of two main DOS centered at $\epsilon_{Li}$ and $\epsilon_{Be}$,
respectively.  For the concentration range under consideration Be states form
the ``majority'' (main) sub-band. As the concentration
$c$ is reduced the impurity sub-band is split off from the main
sub-band (Fig.~\ref{cpa350} (a)). The same effect is seen also when the average DOS
is computed from the alloys components provided by the
TMT effective medium (Fig.~\ref{cpa350} (b)).

In our self-consistent framework  the total number of electrons is
computed from the average DOS $\rho(E)$ of the alloy-components. Therefore the
Fermi level calculated within the CPA is different from the value obtained when
the typical medium is used. Within the CPA, Fig.~\ref{cpa350}(a), the
Fermi level is pinned by the maximum of the impurity band, while
using the typical medium, Fig.~\ref{cpa350}(b), the Fermi level lies between the
main and the split-off band, and a pseudogap
develops around $E_F$. We note that due to the charge
redistribution no split band can be seen in the average DOS
calculated with the TMT. Instead, the shift accounts for the
conservation of the total number of electrons. For all
concentrations  studied here the width of the sub-bands are found to be reduced compared to the CPA results, which already indicates the appearance of localized states at the
band-edges.

\begin{figure}[h]
\includegraphics[width=\columnwidth,clip=true]{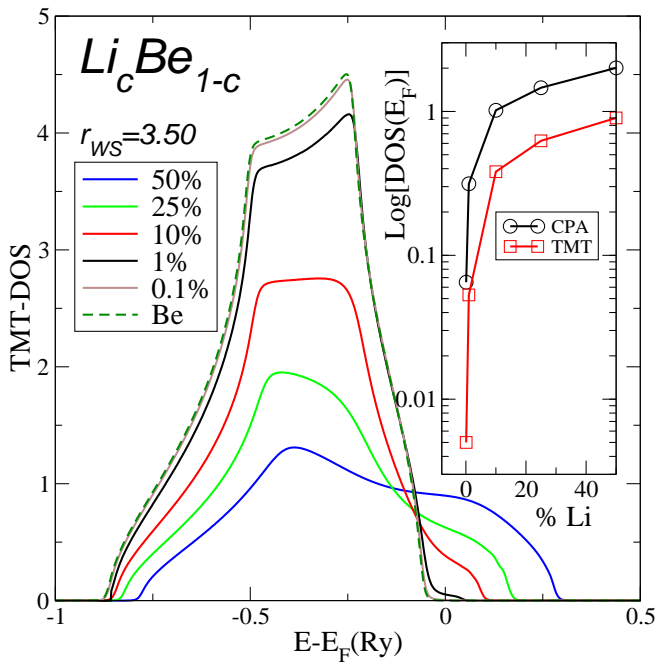}
\caption{The typical medium DOS (order parameter)
of the Li$_{c}$Be$_{1-c}$ alloy for different concentrations $c$ of Li
at the same value of $r_{WS}=3.50$~a.u. (same bandwidth).} \label{tm350}
\end{figure}

An important quantity monitoring the effects of Anderson localization~\cite{sc.sc.10,do.pa.03,dobr.10} is the DOS obtained by the TMT (``TMT-DOS'')~\cite{sc.sc.10,do.pa.03,dobr.10}. It corresponds to an order parameter of the metal-to-insulator transition and
is calculated according to Eq.~(\ref{rho_tm2}) as
\begin{equation}
\rho(E)=\left[ \rho_{Li}(E) \right]^{c_{Li}} \cdot \left[ \rho_{Be}(E)\right]^{c_{Be}}.
\end{equation}

Results are shown in Fig.~\ref{tm350}.
By decreasing the concentration of Li the bandwidth is reduced due to Anderson localization. However, even for $1\%$ of Li the TMT-DOS
at $E_F$ does not reach zero. In the inset
of Fig.~\ref{tm350} we present the comparison of the logarithms of CPA-DOS with
that of the TMT-DOS. We see that at a concentration of $0.1 \% $ the TMT-DOS is reduced
by an order of magnitude.
In this low concentration range the shape of the TMT-DOS closely resembles that of
bulk Be below $0.01\%$, see Fig.~\ref{tm350}. Reducing the concentration  further
decreases the value of the TMT-DOS at $E_F$ even more (see inset of Fig.~\ref{tm350}).

\begin{figure}[h]
\includegraphics[width=\columnwidth,clip=true]{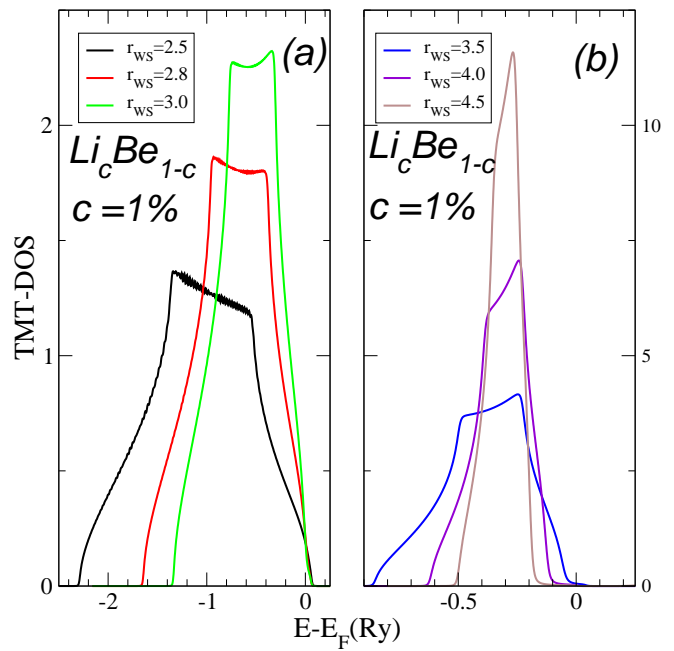}
\caption{The typical medium DOS (order parameter)
of the Li$_{c}$Be$_{1-c}$ alloys for different lattice parameters $r_{WS}$ (different
bandwidths) at the Li concentration $c=1\%$. Note the factor five between the scales
corresponding to the smaller (a) and larger (b) $r_{WS}$. } \label{tm_c}
\end{figure}

In Fig.~\ref{tm_c} we show the dependence of the TMT-DOS on the
Wigner-Seitz radius ($r_{WS}$). Increasing $r_{WS}$ corresponds to a
larger lattice parameter. Expanding the unit cell
leads to narrower bands. For $r_{WS}$ in the range of
$2.5$ to $3.0$~a.u. (Fig.\ref{tm_c}(a)) the TMT-DOS is nonzero at the Fermi level, while a significant
reduction is obtained for even larger values of $r_{WS}$ (Fig.\ref{tm_c}(b)).

\subsection{Magnetism in doped MgO}

The question whether, and how, impurity doping
can stabilize band ferromagnetism has been extensively
studied in the search for novel diluted magnetic semiconductors~\cite{sa.be.10,di.oh.14}.
On general grounds a substitution by an atom with reduced
valency leads to hole-doping which can shift the Fermi level into the
valence band of the insulator or semiconductor. If the shift
is sufficiently strong such that the Stoner criterion is fulfilled,
spontaneous spin polarization sets in. Alternatively, substitution may introduce
spin-polarized impurity states in the gap. When the concentration is increased
the impurity states form bands, which remain spin-polarized. In the following we illustrate this
second scenario in the case of carbon substitution at the oxygen sites
(Sec.~\ref{sec:mgoc}) and manganese substitution at the magnesium sites
(Sec.~\ref{sec:mgmno}) in MgO.
For the sake of completeness, we mention that there are other mechanisms beside
spin-polarization that may lift the degeneracy at the Fermi level, such as polaron formation~\cite{ko.le.18}. However, in the present work we do not
address this mechanism.

MgO is commonly employed as a substrate for thin-film growth of materials
such as metals~\cite{hoel.86}, nitrides~\cite{sc.in.15}, graphene~\cite{ko.bj.10}, and high-$T_c$ superconductors~\cite{mo.ru.90}.
It also finds use as a barrier in magnetic tunnel junctions~\cite{pa.ka.04,ik.ha.08}. This makes MgO an important material
for nanotechnologies and spintronic applications.
Deviations from perfect crystallinity in MgO can have detrimental effects
on its functionality. It was shown that poor-quality MgO substrates gives
rise to poor-quality thin films~\cite{sc.in.15}, while defects and
impurities can lead to a degradation of the superconducting properties in
high-$T_c$ superconductors~\cite{mo.ru.90,ka.sa.06}.
On the other hand it is known that doping Fe/MgO/Fe tunnel junctions with carbon impurities
increases the output voltage and reduces noise~\cite{ti.si.06,he.gu.12}.

Under ambient conditions MgO crystallizes in the rocksalt (B1) structure
with a measured lattice parameter $a_{\textrm{exp}} = 4.216$~{\AA},
and has a large direct electronic band gap of about $7.8$~eV~\cite{ro.wa.67,wi.ar.67,ko.mc.77}.
In the following calculation the experimental lattice parameter is used.
The O-$2s$ and O-$2p$ states as well as the Mg-$3s$ states are treated
as valence states. In a simple picture the two valence
$s$-electrons of Mg fill the O-$2p$ shell.
The bottom of the conduction band
has almost pure Mg-$3s$ character
at the $\Gamma$-point, where the $p$-bands reach
their maximum, making MgO a direct-gap insulator.
While the local density approximation (LDA) is known to underestimate the gap of insulators
and semiconductors~\cite{pe.le.83}, we obtain a direct band gap of $4.50$~eV
which is consistent with values from other
investigations using similar approximations~\cite{ka.st.97}.

The EMTO method was successfully used to study the elastic and
magnetic properties of pure and Fe-doped MgO~\cite{ko.vi.07}.
We now explore other doping effects.

\subsubsection{Carbon-doped MgO (MgO$_{1-c}$C$_c$)}
\label{sec:mgoc}

\begin{figure}[t]
\centering
\includegraphics[width=\columnwidth,clip=true]{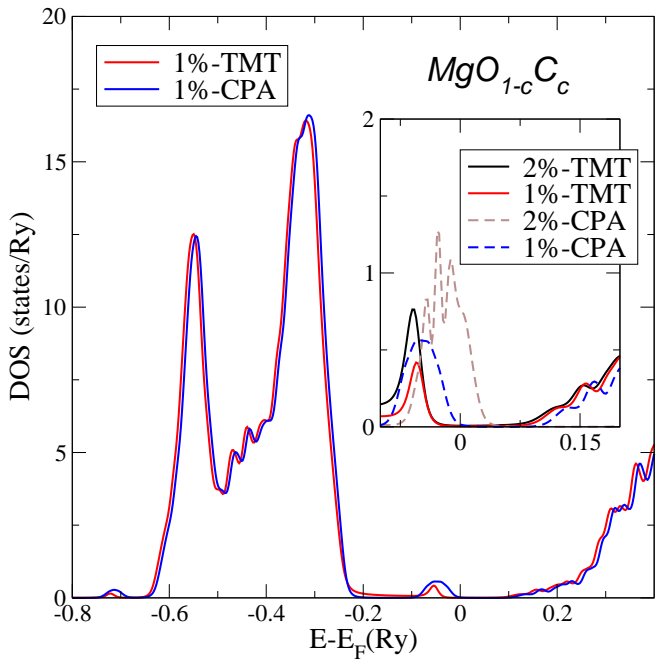}
\caption{Total DOS of MgO$_{1-c}$C$_c$
for $c=1\%, 2\%$C. In the inset the TMT-DOS is compared with the DOS obtained using the effective
medium of the CPA.\label{mgo_dos}}
\end{figure}

The electronic and magnetic properties of vacancies and
non-magnetic impurities in MgO have been investigated in detail for large impurity concentrations~\cite{wu.st.10,sl.ma.11}.
In particular, the polarization of the valence states of MgO by the substitution of oxygen with
$p$-type impurities such as C and N~\cite{el.ru.07,wu.st.10} has been widely discussed. The typical experiments
involve a high concentration of impurities. Consequently
the ferromagnetic interaction is mediated by partially
occupied spin-polarized impurity states. In other words,  the
impurity states should be extended to mediate ferromagnetism,
and at the same time the band width should be small enough
such that the Stoner criterion is satisfied~\cite{pa.pi.08,pe.xi.09}.
We note that in LDA and the generalized gradient approximation (GGA)
the impurity $2p$-states are too extended and the
magnetism is overestimated for high concentrations~\cite{pa.pi.85,ch.la.09}.
Attempts to improve these results by including a Hubbard $U$ within a Hartree mean-field decoupling
(LDA$+U$ or GGA$+U$ approximations) or, alternatively, by including self-interaction corrections produce a splitting of the $2p$-impurity levels and thereby lead to ferromagnetism. However, the local distortions around the defects due to the relaxation of the crystal structure inhibit magnetism~\cite{dr.pe.08,ch.la.09}.
A different mechanism for the formation of a magnetically ordered state, valid in the high concentration regime, is the formation of
impurity pairs~\cite{wu.st.10}.

We now explore the opposite limit of very low
concentrations ($0.01\% \le c\le 2\%$) when impurity states are well isolated.
Then the question arises whether ferromagnetism is also found when localization effects are
taken into account through the TMT approach.
In Fig.~\ref{mgo_dos} we show the DOS of MgO
with $1\%$ C impurities calculated for the CPA and TMT
effective medium, respectively. Using the CPA effective medium  a non-vanishing value of the DOS
at $E_F$ is obtained. By contrast, a significantly
reduced value of the DOS at the Fermi level is found in the TMT.
In the inset of Fig.~\ref{mgo_dos} the order parameter (TMT-DOS) is compared with the
CPA-DOS for $c=1\%$ and  $2\%$ C. At both  concentrations the
TMT-DOS is already zero, indicating an
insulating state.
At higher carbon concentrations MgO
has a magnetic ground state~\cite{wu.st.10}.
We notice that already for $2\%$ carbon doping the ferromagnetic phase is lower in the total energy
than the non-magnetic one.

\begin{figure}[h]
\centering
\includegraphics[width=\columnwidth,clip=true]{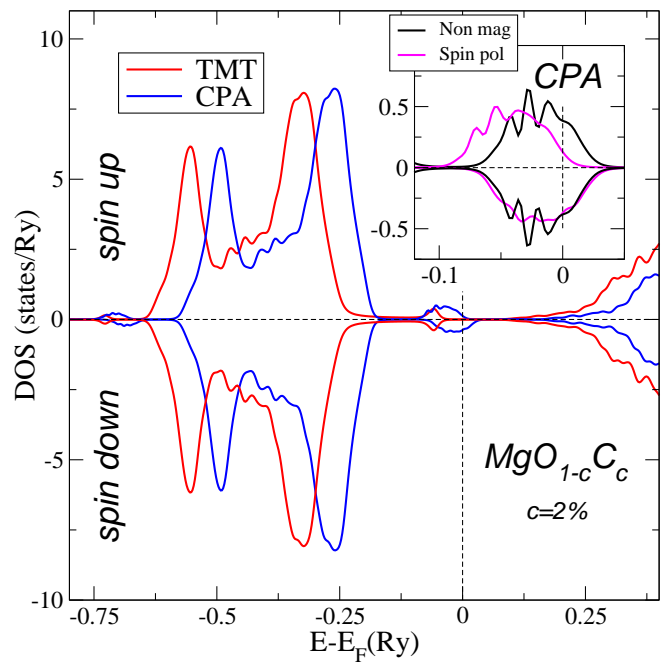}
\caption{Total DOS obtained with the spin-polarized CPA  for $2\%$ C substitution
in MgO$_{0.98}$C$_{0.02}$ (blue line) and the non-magnetic TMT
solution (red line). Inset: comparison of magnetic and non-magnetic CPA
solutions.} \label{fig:mag-nonmag}
\end{figure}

In Fig.~\ref{fig:mag-nonmag} we show the DOS for MgO$_{0.98}$C$_{0.02}$ in the
ferromagnetic phase computed within CPA together with the non-magnetic solution obtained by the TMT.
The inset of Fig.~\ref{fig:mag-nonmag} shows a comparison of the CPA results in the non-magnetic and ferromagnetic phase, respectively.
In the ferromagnetic state  LDA predicts that the carbon acquires
a non-zero magnetic moment of about $0.55$~$\mu_B$, which increases in GGA
to about $0.95~\mu_B$. At $2.5 \%$C the GGA produces an almost totally spin-polarized state
with spin-up DOS merging the main band. By contrast, for the typical medium a stable magnetic solution was neither found in  LDA or GGA.

\subsubsection{Manganese-doped MgO  (Mg$_{1-c}$Mn$_c$O)}
\label{sec:mgmno}

A half-metallic state was predicted for Mn-doped MgO, with Mn replacing Mg~\cite{sh.pi.11}. Both a high-spin ($S=5/2$) and
a low-spin ($S=1/2$) state can be stabilized, with the high-spin state being insulating and the low-spin state
being half-metallic~\cite{me.bo.14}. The high-spin state is relevant for quantum information science since the system
has been used to host qubits~\cite{be.ch.09}.

\begin{figure}[h]
\centering
\includegraphics[width=\columnwidth,clip=true]{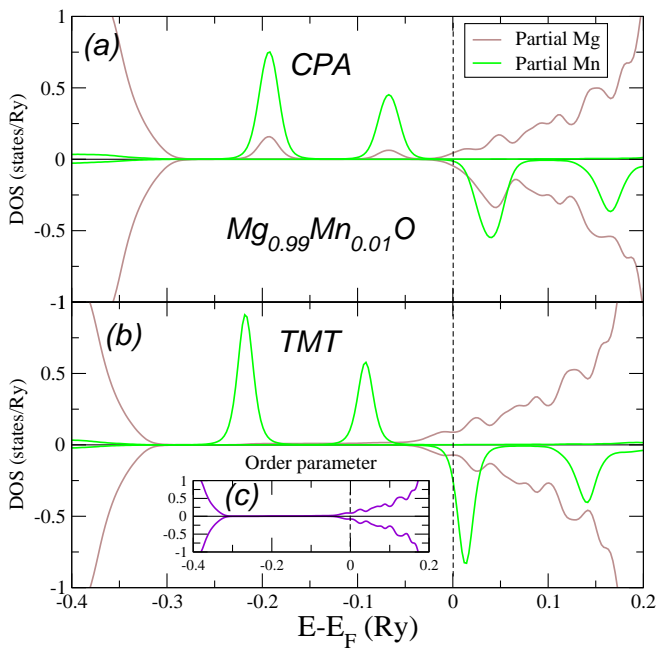}
\caption{(a) Spin-resolved CPA-DOS in the high-spin state for Mn-doped MgO.
Brown/green curves correspond to Mg/Mn partial DOS, respectively. (b)
TMT spin resolved partial DOS. Inset (c): geometric average of the
partial DOS (order parameter).} \label{fig:highspin}
\end{figure}

Doping with $d$-electrons raises the question regarding the influence of $p-d$ hybridization upon the magnetism in the
Mg$_{1-c}$Mn$_c$O alloy. On general grounds the $p$ and the $d$-bands are coupled by the hybridization and
disorder effects through the effective medium potential.
Quite generally the hybridization determines the relative band shifts,
while disorder broadens the shifted bands.
Magnetism depends on the $d$-occupations. Due to the disorder
the fluctuation of the number of particles  will influence the propagation of electrons and thus change the DOS. Since the fluctuation is spin--dependent it affects the spin up and down
DOS differently and thereby directly influences the spin asymmetry.

In Fig.~\ref{fig:highspin}(a) the spin-resolved DOS for MgO doped with $1\%$ Mn as computed with CPA (upper panel)
and TMT (lower panel), is shown. The Mn $d$-states are located in the gap of MgO and are split according to the
cubic crystal field: the triply degenerate $t_{2g}$ states are located at about $-0.2$~Ry and the doubly degenerate $e_g$ states
at about $-0.1$~Ry. In this case the ground state  corresponds to the high-spin solution with both spin-up
$t_{2g}$ and $e_g$ states being completely filled, while the spin-down $t_{2g}$ and $e_g$ states are shifted above the
Fermi level. Note that the partial Mg and Mn DOS obtained within the CPA shows overlapping maxima
at the same energies as a consequence of significant $p-d$ hybridization. The position of the maxima corresponds to the
shifted one-particle energies because of hybridization. The value of the Mn magnetic moment is $3.8~\mu_B$ and
corresponds to the $S=5/2$ state.
By contrast the DOS obtained within the TMT, Fig.~\ref{fig:highspin}(b), shows no overlap of $p$ and $d$ orbitals, and
the total DOS within the gap is determined only by Mn $d$-orbitals. The absence of hybridization leads to a shift
of the $t_{2g}$ and $e_g$ states. The Mn total magnetic moment is $3.8~\mu_B$.
The inset
of Fig.~\ref{fig:highspin}(b) shows the geometric average of the LDOS (at the Mg-site), which
represents the order parameter. On the level of the DOS the self-consistent TMT effective medium
 calculation leads to a cancellation of the $p-d$ hybridization.

\begin{figure}[h]
\centering
\includegraphics[width=\columnwidth,clip=true]{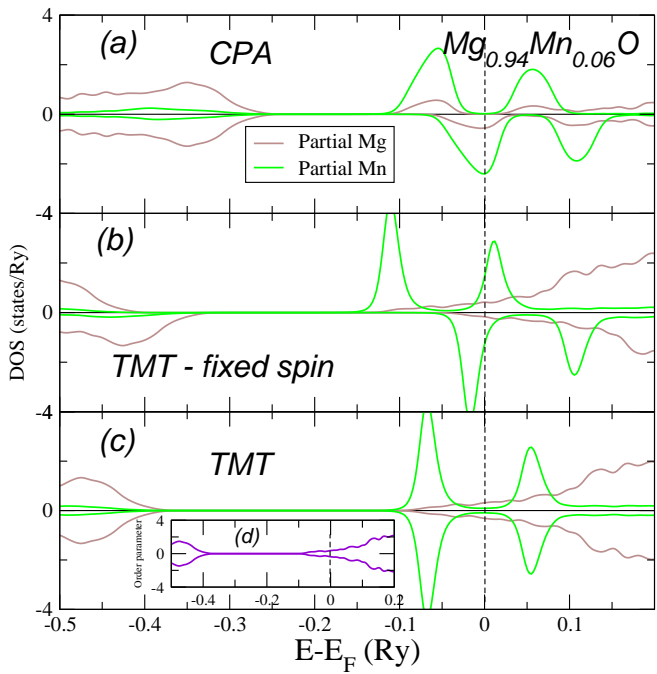}
\caption{(a) Spin-resolved CPA-DOS in the low-spin state.
Brown/green curves correspond to Mg/Mn partial DOS, respectively.
Fixed-spin(b) and un-biased spin polarized calculation (c) using the TMT. Inset (d): geometric average of
partial DOS (order parameter).
} \label{fig:lowspin}
\end{figure}

Fig.~\ref{fig:lowspin}(a) shows the CPA solution for the low-spin state in the case of $6\%$ Mn doping.
Here the majority $t_{2g}$ states are completely filled and the minority $t_{2g}$ states are partially occupied.
The $e_g$ states are empty in both spin channels. The resulting state is half-metallic with a spin $S=1/2$.
No low-spin magnetic configuration is found within the TMT. Instead, by fixing
the total spin to the expected low-spin solution ($0.06~\mu_B$), the TMT converges to a state with a DOS shown
in Fig.~\ref{fig:lowspin}(b). Since the fixed-spin solution was allowed to relax,
the system converges to a non-magnetic solution (see Fig.~\ref{fig:lowspin}(c)).
In both TMT computations
Figs.~\ref{fig:lowspin}(b) and (c) no overlap between the $p$ and $d$
orbitals is found.

We now summarize the comparison between the CPA and TMT results in the case of Mg$_{1-c}$Mn$_c$O.
The question if, and how, the $p$-$d$ hybridization contributes to the magnetic properties of this alloy
can indeed be answered. In the case of CPA the $p$-$d$ hybridization is present and both the low- and high-spin magnetic configurations are possible.
However, within the TMT the high-spin state remains magnetic with no significant difference compared with the corresponding
CPA solution, despite the suppressed $p$-$d$ hybridization. Furthermore, no low-spin magnetic configuration is obtained
using the TMT effective medium. The destabilization of the low-spin solution results from the spin-dependent particle fluctuations between $p$ and $d$ states, due to disorder. In fact, both CPA and TMT have such effects included
through the effective medium `self-energy' which renormalizes the quasiparticle DOS. This
renormalization can be analyzed in terms of band broadening and band shifts. The inter-band fluctuations are
found to have considerable influence on the magnetic properties of the low-spin configuration. Namely,
these weak magnetic effects remain for weak disorder (CPA), but are suppressed by strong disorder (TMT).

\section{Conclusion}

We implemented the effective typical
medium theory within the DFT by employing the
EMTO-basis set and compared the results obtained thereby  with those of the CPA.
The framework was then applied to study the evolution of the impurity
band appearing in the hypothetical Li$_{c}$Be$_{1-c}$ alloy, a  simple cubic system with one-atom per unit cell. This  alloy
system was studied within DFT and different effective medium theories.
The DOS results show signatures of Anderson localization effects, band narrowing
and split-off impurity bands.
We also discussed the effect of charge self-consistency and the
determination of the Fermi level $E_F$. The search for $E_F$ is implemented
in a similar way as in the CPA case, with the difference that the
average DOS is constructed from the alloy components computed with the
typical medium path operators. Furthermore, we studied  the magnetic properties
of dilute MgO$_{1-c}$C$_{c}$ and Mg$_{1-c}$Mn$_c$O alloys. In contrast to
the carbon substitution in MgO the Mn substitution brings into the discussion the
presence of $d$-orbitals and their contribution to the magnetism of
Mn impurities. We found that disorder-induced interband particle number
fluctuations suppress magnetism together with the $p$-$d$ hybridization in the
case of low-spin configurations. In the charge self-consistent calculations
the system lowers its energy by particle fluctuations. Since these fluctuations are
spin dependent they influence the  magnetic stability of the system.

The implementation of the effective typical
medium theory within the DFT presented in this paper can be generalized to include electronic correlation effects
through the dynamical mean-field theory (DMFT). This opens the possibility for
\emph{ab initio} studies of correlated electron materials in the presence of disorder beyond CPA.

\subsection*{Acknowledgments}
This work was supported by the U.S. Department of Energy, Office of Science, 
Office of Basic Energy Sciences under Award Number DE-SC0017861.
Financial support offered by the Augsburg Center for Innovative Technologies,
and by the Deutsche Forschungsgemeinschaft (DFG), Projektnummer 
107745057 - TRR 80/F6, is gratefully acknowledged.
HT gratefully acknowledges support from NSF grant CSSI-1931367.
LV acknowledges financial support from the Swedish Research Council,
the Swedish Foundation for Strategic Research, the Swedish
Foundation for International Cooperation in Research and
Higher Education, and the Hungarian Scientific Research Fund (OTKA 84078).
We acknowledge computational resources provided by the Swedish National
Infrastructure for Computing (SNIC) at the National Supercomputer Centre (NSC) in Link\"{o}ping.

\bibliography{references,new_bib}

\end{document}